\documentclass[twocolumn]{article}

\usepackage{amsmath,amsfonts,epsfig,graphics,amssymb}
\usepackage{authblk}
\usepackage[margin=1.8cm, top = 2.2cm, bottom=3cm]{geometry}

\usepackage{hyperref}
\hypersetup{colorlinks=true,citecolor=blue}

\title{\bf Analysis of published data of electron capture in $^7$Be in the search for a heavy neutrino in the mass range under 800~keV}
\author{N.\,A.~Likhovid}
\author{V.\,S.\,Pantuev}

\affil{\small \it Institute for Nuclear Research RAS, 117312 Moscow, Russia}

\date{} 
\begin{document}
\twocolumn[
\begin{@twocolumnfalse}
\maketitle
\begin{abstract}{We present reanalysis of the experimental data of electron capture in $^7$Be embedded in Ta which have been published by other authors. Our goal is to set upper limits on a mixture of electron neutrino with a possible right handed heavy neutrino in the 150--800~keV mass range. In the published  experiment  a $^7$Li recoil energy spectrum in the 20--200~eV range was measured. In case of electron capture with emission of a heavy neutrino, the recoil spectrum should be shifted to the lower energies. We search for an additional Gauss-shaped structure with the same energy width as the main K-shell  transition peak. For this we digitize the published spectrum curve, find the energy resolution, calculate the moving sum of the events along the spectrum in the energy interval of about 3 sigma of energy resolution. Then we use the statistical error of this sum to exclude at some level the appearance of an  additional peak. Finally, we present the upper limits at a 95\% confidence level on electron neutrino -- heavy neutrino mixing element, $U^2$, in the mass matrix. New upper limits are at least one order of magnitude lower than the existing data in 300--800~keV mass range.
}
\vspace{\baselineskip}
\end{abstract}
\end{@twocolumnfalse}
]

\vspace{\baselineskip}

\paragraph{1. Introduction.}
 Neutrinos are massive and this property cannot be accommodated in the Standard Model. The simplest mechanism to provide neutrino with  mass assumes the existence of new particles - right handed (or "sterile") neutrinos. Their number, masses and mixing angles with left handed ("active") neutrinos, apart from the  existing observational restrictions, are free parameters of the theory~\cite{Adhikari:2016bei}.  Sterile neutrino in the keV mass range is one of the best motivated dark matter particle candidates. Therefore, studies of the sterile neutrino mass may probe a new physics. Currently there have been many attempts to search for such heavy neutrino  directly in the laboratory~\cite{Adhikari:2016bei}. Nucleus $\beta$-decay and electron $K$-capture are prominent channels to search for a heavy neutrino component~\cite{Holzschuh:2000nj}, ~\cite{Schreckenbach:1983cg}. 
 
 Electron capture in $^7$Be is a two body decay with emitted neutrino and recoil $^7$Li nucleus. Almost 20 year ago  micro calorimeters were used to detect recoil $^7$Li nucleus where $^7$Be was implanted into HgTe chips~\cite{Voytas:2002zz}.  High statistics data for electron capture from $K$- or $L$-shells in $^7$Be have recently been published~\cite{Fretwell:2020ntq}. In this paper the authors used Ta-based superconducting tunnel junction (STJ) detectors to measure a recoil nucleus in the 10--200~eV energy range. We decided to use these data for estimation  of the possible admixture of a heavy neutrino in contrast to emission of the standard electron neutrino. After some steps taken to re-treat the published spectra, we extract upper the 95\% CL limits for a such heavy neutrino in the mass range 150--800 keV.
\paragraph{2. Features of $^7$Be decay.}
$^7$Be decays by electron capture from $K$- or $L$- orbit to $^7$Li ground or first excited state with $Q$ value of about 862~keV. 
Thus, it has four possible channels: 
\begin{itemize}
	\item  with a 89.5\% probability the decay is to $^7$Li ground state and a neutrino with the total energy of 862 keV is emitted and the residual recoil nucleus gets 57~eV kinetic energy;
	\item  with a 10.5\% probability the decay may go to  $^7$Li excited state and the neutrino  total energy is 384 keV. This decay is quickly followed by a wide angle gamma emission with energy of 478 keV. The recoil nucleus energy spectrum becomes wide;
	\item if the capture is from $K$-shell an Auger electron is emitted from $L$-shell and its energy of about 55 eV adds to decay signal with a total energy of 112~eV. It allows one to separate $K$- and  $L$ -capture peaks.
		\item the capture from $L$-shell to the exited state is followed by gamma emission.
\end{itemize}
As a result, the recoil nucleus spectrum has two similar parts: a prominent $K$ line peak at 112 eV  with a wide distribution from 55 eV to 112 eV centered at about 80 eV when the decay goes to the excited state, and $L$ line peak at 57 eV  and a relevant wide distribution centered  at about 30 eV for decays to the excited state. 
These features were clearly seen in papers~\cite{Voytas:2002zz} and~\cite{Fretwell:2020ntq}. 

The letter paper has an additional feature: $K$- line peak is not at 112 eV but roughly at 107 eV. The authors of paper~\cite{Fretwell:2020ntq}  consider a list of in-medium effects to describe the whole energy spectrum including electron escape and $shake-up$ and $shake-off$ in  daughter $^7$Li atom (see the paper for all explanations). They also state that the observed difference could be due to in-medium effects of Li in Ta.  Indeed, this shift of about 4-5 eV could be attributed to the difference in $^7$Be source preparation and detector materials. As already mentioned, the additional 55 eV should come from the Auger electron energy.  According to the results and calculations in Ref.~\cite{hovington} the intensity and energy of this transmission depend on chemical bonding of $^7$Li. 

\paragraph{3. Procedure.}
One of the main motivations for our work comes from the plot of residuals after applying calibration and describing the spectrum distribution~\cite{Fretwell:2020ntq}. Recoil spectrum of  $^7$Li does not have any statistically significant features and the single channel residuals in most cases do not exceed two sigmas. Thus, our task was to transfer a high statistics spectrum to estimation of the upper level for a possible heavy neutrino admixture. 

\begin{figure}[htb]
	\begin{center}
		\includegraphics[width=.95\linewidth]{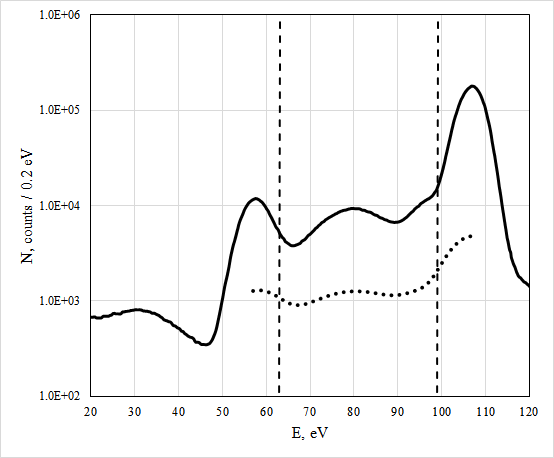}
	\end{center}
	\caption{Figure 1. Solid line -- our reconstructed energy spectrum from Ref.~~\cite{Fretwell:2020ntq}; dotted line -- double statistical error in the moving interval in search for a sign of heavy neutrino, see text. In the  region between two vertical dashed lines a such statistical search is still valid.}
	\label{fig:spectrum}
\end{figure}

We assume that the detector response can be described by a finite resolution with Gaussian shape, the width of which does not depend on recoil energy. Formally, an additional heavy neutrino component should demonstrates a similar recoil energy spectrum but shifted to the lower range. Thus, first of all, we should search for other $K$- and $L$- line peaks to the left of the main ones.
Our analysis has the following steps:
\begin{itemize}
	\item  digitization of the published recoil spectrum lines from the plot, we fill our own histogram with the same bin size of 0.2~eV, solid line in Fig.~\ref{fig:spectrum}. We estimate systematic error from a such digitization to be not more than 5\%, which is defined by 1-2 pixel size in the original figure, and was neglected;
	\item estimation of energy resolution by fitting by Gauss function the main $K$-line; we get $\sigma$= 3.0~eV or 15 bins;
	\item calculation along the spectrum of the moving sum  over 0.2~eV bins for the nearest  intervals of about 3$\sigma$ or 9~eV width. Such a width is optimal for a flat background.  The actual interval width was slightly varying to get the optimal ratio between the potential signal and the underlying  background with changing shape. For this, we calculate control values  from each reference bin at energy $E_B$ to the left edge, 
$E_L, L_{erf}=erf[(E_B-E_L)/\sqrt{2}\sigma]$, and to the right edge, $E_R, R_{erf}=erf[(E_R-E_B)/\sqrt{2}\sigma]$, $erf()$ is the error function,  and try to minimize control value  
$Cr=4\sqrt{Sum}/(L_{erf}+R_{erf})$, where $Sum$ is a sum of the events from the left to the right edge of the interval;	
	\item calculation of statistical error for each interval by taking square root from the sum;
 	\item correction of each sum for potential missing counts in Gauss tails beyond the  interval; 
	\item estimation of a 95\% confidence level (CL) multiplying the statistical error by 1.95, dotted line in Fig.~\ref{fig:spectrum}. This value is used to set a 95\% upper CL;
	\item correction of probability to decay to excited state dividing by its probability of 0.895. The reason is that we do not search for heavy neutrino with a wide recoil energy distribution for the case to decay to $^7$Li excited state;  
	\item calculation of CL by dividing corrected value by the number of electron capture from $K$-shell;
		\item finally, we correlate the recoil energy  and relevant 95\% CL with the mass of a possible heavy neutrino.
\end{itemize}
We estimate a possible systematic error by assumption that the energy resolution or the shape of the peak are know with 10\% uncertainty, say, with would be $\pm$0.3~eV for  sigma of the Gauss peak. This systematic translates to about 5\% of CL.

The procedure described above, actually, gives  the sensitivity limit based on  the accumulated statistics. In a real experiment, the search for a small peak will meet an additional problem: it is difficult to find a such peak in vicinity to a large Gauss-like one. Some procedure or peak-finding algorithm should be applied. In Fig.~\ref{fig:spectrum}  we plot two vertical dashed lines which define the central region between $L$ and $K$ peaks where  the sensitivity limit estimation is valid. For energy regions at about 57--63~eV and 99--107~eV we did an additional simulation to find the actual limits. To do this, we:
\begin{itemize}
	\item simulate a small Gauss-like peak on shoulder of the large peak. The width of the simulated peak is equal to  the energy resolution or 3~eV;
	 \item sum two distributions;
	 \item fit this sum by a single normal distribution function; 
	 \item calculate residuals between the sum and the fit;
	  \item compare the total absolute value of the residuals with an estimation of statistical fluctuation in this region; 
	   \item at particular position of the small peak find its minimal recognizable magnitude.
\end{itemize}
This gives an additional conservative confidence level for a small peak in vicinity to the large one. 

\begin{figure}[ht]
	\begin{center}
		\includegraphics[width=.95\linewidth]{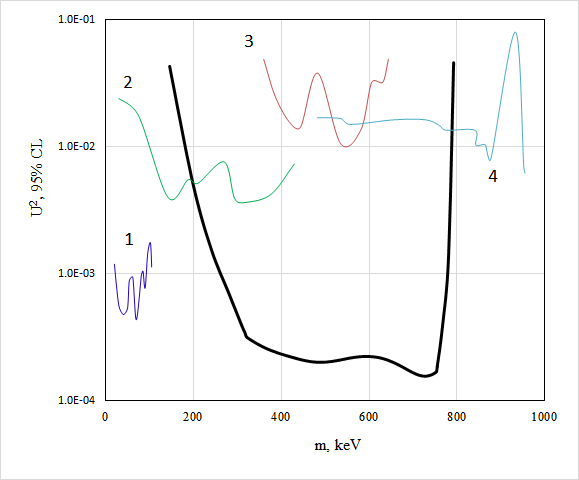}
	\end{center}
	\caption{Figure 2.  Upper 95\% CL limits, $U^2$, for probability  to find a heavy neutrino versus its mass. Thin solid numbered lines are for published data: 1 -- ~\cite{Holzschuh:2000nj}, 2 -- ~\cite{Schreckenbach:1983cg}, 3 -- ~\cite{3}, 4 -- ~\cite{4}.} 
	\label{fig:limits}
\end{figure}

\paragraph{4. Results.}
In Fig.~\ref{fig:spectrum} we plot the reconstructed energy spectrum released by a $^7$Li recoil atom, solid curve. 
 The plot of the residuals in~\cite{Fretwell:2020ntq} does not have any statistically significant features, thus, as already mentioned,  we do not actually search for an additional peak, but made a statistical analysis of the spectrum to exclude any possible additional component. The dotted line in Fig.~\ref{fig:spectrum} represents a double statistical error (actually, 1.95 error to get later a 95\% confidence level) for the moving sum over about 3 sigma energy interval. The results were extracted from $K$-line data between 55 eV and 107 eV. 
For electron capture from the $K$-shell to the $^7$Li ground state the neutrino with zero mass produces a peak at 107 eV in the measured spectrum. The largest possible heavy neutrino mass of 862 keV should form events around energy of Auger electron or close to 50 ~eV. Thus, calculating relation between the energy positions of the dotted line in Fig.~\ref{fig:spectrum}  and the possible heavy neutrino mass and taking into account the statistical significance , we get  upper 95\% CL limits, $U^2$, for probability to find a heavy neutrino versus its mass,   Fig.~\ref{fig:limits}, solid line. Limits at the regions  150--300~keV and at 750--800~keV are obtained by simulation close to $L-$ and $K-$peaks/ as described. At the same figure we plot  all existing published data. New limits are at least one order of magnitude lower.

To conclude, we present our estimation of limits on search for a heavy right-handed (sterile) neutrino. We use the published high statistics data from  reference~\cite{Fretwell:2020ntq} where  they measured electron capture in $^7$Be. 
Electron capture in $^7$Be is a two body decay with emitted neutrino and recoil $^7$Li nucleus. Neutrino mass defines the maximum energy released by $^7$Li. After performing statistical analysis of the measured spectrum we get an upper 95\% CL  in the mass range 150--800 keV. These limits are at least one order of magnitude lower than the existing published data.

 We would like to thank our colleagues A.~Nozik, D.~Abdurashitov and L.~Inzhechik  for useful discussions.
 
 This work is supported by the Ministry of Science and Higher Education of
the  Russian Federation under the contract 075-15-2020-778.

\end{document}